\documentclass[12pt]{article}
\usepackage{epsfig}
\usepackage{amssymb}
\usepackage{pslatex}

\newcommand{\bea}{\begin{eqnarray}}
\newcommand{\eea}{\end{eqnarray}}
\newcommand{\be}{\begin{equation}}
\newcommand{\ee}{\end{equation}}
\def\rmd{{\rm d}}

\begin{document}

\vfill \setcounter{page}{0} \setcounter{footnote}{0}
\begin{titlepage}
\hfill

\vspace*{15mm}

\begin{center}
{\large {\bf Domain Walls in AdS-Einstein-scalar Gravity}\\
}
\vspace*{18mm} {Sangheon Yun}\\
{ \vspace{5mm} Institute for the Early Universe, Ewha Womans University,\\
Daehyun-dong, Seodaemun-gu, Seoul 120-750,~ KOREA \\
\vspace*{29mm} }

\end{center}

\begin{abstract}
In this note, we show that the supergravity theory which is dual to ABJM field theory can be consistently reduced to scalar-coupled AdS-Einstein gravity and then consider the reflection symmetric domain wall and its small fluctuation. It is also shown that this domain wall solution is none other than dimensional reduction of M2-brane configuration.
\end{abstract}

\vspace*{30mm}
\vspace*{10mm}

E-mail address : sanhan@ewha.ac.kr
\end{titlepage}

\section{Introduction}

In \cite{Aharony:2008ug} a new progress in AdS/CFT correspondence\cite{Maldacena:1997re} was found: the ABJM field theory, i.e. three-dimensional $\mathcal{N} = 6$ superconformal $U(N)\times U(N)$ Chern-Simons theory with Chern-Simons level $(k,-k)$, which was conjectured to describe the low energy dynamics of a stack of $N$ coincident $M2$-branes probing the singularity of a $\mathbb{C}^4 / \mathbb{Z}_k$ orbifold background, is holographically dual to the type IIA string theory on $AdS_4\times \mathbb{CP}_3$ or to M-theory on $AdS_4\times S^7 / \mathbb{Z}_k$.
After this discovery, there has been remarkable progress in understanding AdS$_4$/CFT$_3$ correspondence.
It is based on the large $N$ planar limit where one is taking $N,\  k\rightarrow \infty$ while holding the 't Hooft coupling $\lambda={N\over k}$ to be fixed and large.
The ABJM theory has $SU(4)$ R-symmetry, which corresponds to the symmetry of $\mathbb{CP}_3$ whose isometry $SU(4)$ preserves not only the metric, but also the Kahler form.
The theory has only $\mathcal{N} = 6$ manifest supersymmetry, but is superconformal, and has bifundamental matter and a specific quartic superpotential.
The inverse Chern-Simons level $1/k$ plays a similar role to the Yang Mills coupling constant $g^{2}_{YM}$.
If $\lambda$ is small, the superconformal Chern-Simons theory is weakly coupled and it can be studied by perturbative analysis.
On the other hand, for strongly coupled regime of large $\lambda$, the supergravity is a valid description.
There are still many outstanding questions, which would be valuable for research implementation in the future, and one of them is to investigate thermal aspects of the ABJM model in both the strongly and the weakly coupled regimes\cite{Bak:2010yd,Smedback:2010ji,Bak:2010qb}.
In studying the low energy dynamics, the compactification of the type IIA supergravity down to four dimensions is very useful.
Using the resulting four-dimensional action, we may consistently study full nonlinear effects.
Therefore, in this note we will first summarize the dimensional reduction of the type IIA supergravity on ABJM background and then find a solution of the low-energy effective theory other than AdS space, AdS-Schwarzschild and AdS-Reissner-Nordstrom black branes.

This paper is organized as follows.
In section 2 we will briefly review the type IIA supergravity and its compactification on $\mathbb{CP}_3$.
In section 3 we will obtain a domain wall solution of the reduced four-dimensional Einstein gravity coupled to a scalar field and comment on other theories coupled with a few scalars.
Then we will conclude this paper with discussion and future directions in section 4.

\section{Type IIA supergravity and compactification on $\mathbb{CP}_3$}

We can construct the effective action of type IIA supergravity via the Kaluza-Klein mechanism by compactification of the eleven-dimensional $\mathcal{N}=1$ supergravity on $S^1$ and then retaining only the massless sector.
Type IIA supergravity can be also obtained from the zero-slope limit of type IIA superstring theory.

\subsection{Type IIA supergravity}

The bosonic part of the action of ten-dimensional type IIA supergravity is written in the string frame,
\be
S^s_{IIA}= S^s_{NS}+ S^s_R+ S_{CS}\,,
\ee
where
\bea
S^s_{NS}&=& \frac{1}{2\kappa^2_{10}}\int d^{10} x \sqrt{-g}\,\,\Big[ \,e^{-2\phi}\bigg( R + 4\nabla_M\phi\nabla^M\phi-\frac{1}{12}H_{MNP}H^{MNP} \bigg)\, \Big]\,, \nonumber\\
S^s_{R}&=& \frac{1}{2\kappa^2_{10}}\int d^{10} x \sqrt{-g}\,\,\Big[ -{1\over 4}\, F_{MN}\,\,F^{MN}-{1\over 48}\, \widetilde{F}_{MNPQ} \,\widetilde{F}^{MNPQ} \Big]\,, \nonumber\\
S_{CS}&=& \frac{1}{2\kappa^2_{10}}\int \,{1\over 2} \,\,B_{[2]}\,\wedge\, F_{[4]}\,\wedge\, F_{[4]}\,.
\eea
Here we have used the following definitions: $F_{[2]}=dA_{[1]}$, $F_{[4]}=dA_{[3]}$.
The NS-NS three form field strength is defined by $H_{[3]}=dB_{[2]}$ and the gauge invariant four form field strength $\widetilde{F}_{[4]}$ by $\widetilde{F}_{[4]}= F_{[4]} + H_{[3]}\wedge A_{[1]}$.
The ten-dimensional gravity coupling in this action is given by $2\kappa^2_{10}= (2\pi)^7 \, \ell_s^8 g_s^2$ where $g_s$ denotes the string coupling.
We define the dilaton field $\phi$ as its nonzero mode by subtracting its constant part $\hat\phi$ denoting the string coupling $g_s=e^{\hat{\phi}}$.

The bosonic part of type IIA supergravity in the Einstein frame is described by the action,
\be
S^E_{IIA}= S^E_{NS}+ S^E_R+ S_{CS}\,,
\ee
where
\bea
S^E_{NS}&=& {1\over 2 \kappa^2_{10}}\int d^{10} x \sqrt{-g}\,\,\Big[ \,\,R(g)-{1\over 2}\nabla_M\phi\nabla^M\phi -{1\over 12}e^{-\phi} \,H_{MNP}H^{MNP}\,\, \Big]\,, \nonumber\\
S^E_{R}&=&{1\over 2 \kappa^2_{10}}\int d^{10} x \sqrt{-g}\,\,\Big[ -{1\over 4}e^{{3\over 2}\phi}\, F_{MN}\,\,F^{MN}-{1\over 48}
e^{{1\over 2}\phi}\, \widetilde{F}_{MNPQ} \,\widetilde{F}^{MNPQ} \Big]\,.
\eea
The string frame metric is given by the transformation $g^{s}_{MN}= e^{{1\over 2}\phi} g^E_{MN}$.
The equations for the bosonic sector of type IIA supergravity can be given in terms of gauge invariant fields as in \cite{Bak:2010yd}.

\subsection{AdS$_4\times \mathbb{CP}_3$ background of ABJM theory}

The near-horizon geometry of $Nk$ supersymmetric M2-branes becomes AdS$_4\times \mathbb{S}^7$ with a radius $R$,
\be
\rmd s^2 = R^2  \left[ {1 \over 4} \rmd s^2(\mbox{AdS}_4) + \rmd s^2 (\mathbb{S}^7) \right]\,,
\ee
in which there are 32 Killing spinors.
In the large $Nk$ limit, the radius $R$ becomes much larger than the eleven-dimensional Planck length $l_P$ and the eleven-dimensional supergravity description works.
The metric has the $SO(2,3)$ conformal symmetry of AdS$_4$ and $SO(8)$ symmetry of $\mathbb{S}^7$.
The unit round seven-sphere $\mathbb{S}^7$ can be presented as a Hopf fibration of a circle over $\mathbb{CP}_3$ base,
\be
\rmd s^2(\mathbb{S}^7) =\rmd s^2(\mathbb{CP}_3) +(\rmd \upsilon+\omega)^2\,,
\ee
where $0\le \upsilon < 2\pi$, and we parametrize the one form $\omega$ explicitly as
\bea
\omega= - {1\over 2} \Big( {\cos 2\xi } \rmd \eta + \cos^2\xi{\cos\theta_1} \rmd \phi_1 +\sin^2\xi{\cos\theta_2 } \rmd \phi_2\Big)\,\,.
\eea
Then $\omega$ gives the K\"ahler two form by $J_{\rm }={1\over 2}d\omega$, which has the following properties:
\be
\nabla_a\, J_{bc}=0\,,\,\,\,\,\,J_{ab}\, J^{bc}=-\delta_{\,a}^{\,c}\,,\,\,\,\,\,8 J^{ef}=\epsilon^{abcdef}\,J_{ab}\,J_{cd}\,,
\ee
with the indices raised by unit $\mathbb{CP}_3$ metric.
If we perform the $\mathbb{Z}_k$ orbifold of $\mathbb{S}^7$, the $SO(8)$ symmetry is broken to $SU(4)\times U(1)$ but the conformal invariance remains.
This is the bosonic symmetry of the ABJM theory.
On the other hand, 24 out of 32 Killing spinors survive in AdS$_4\times \mathbb{S}^7/\mathbb{Z}_k$ with $k>2$.
In the large $Nk$ limit, the $\mathbb{CP}_3$ of the near-horizon metric
\bea
\rmd s^2 &=&R^2  \left[ {1 \over 4} \rmd s^2 (\mbox{AdS}_4) + \rmd s^2 (\mathbb{S}^7/\mathbb{Z}_k) \right] \nonumber\\
&=&R^2  \left[ {1 \over 4} \rmd s^2 (\mbox{AdS}_4) + \rmd s^2 (\mathbb{CP}_3) + \bigg(\frac{\rmd\upsilon}{k}+\omega\bigg)^2 \right]
\eea
remains large because $R\sim (Nk)^{1/6}$.
In the $N\rightarrow\infty$ limit with $N/k$ fixed, the compactification radius $R_{11}=R/k$ along the $\upsilon$ direction becomes very small (compared to the eleven-dimensional Planck length $l_P$), and the metric is reduced to AdS$_4\times \mathbb{CP}_3$.

The supergravity background of the ABJM theory is given by \cite{Aharony:2008ug,Nilsson:1984bj}
\bea
\rmd s^2 &=&R^2_s  \left[ {1 \over 4} \rmd s^2(\mbox{AdS}_4) + \rmd s^2 (\mathbb{CP}_3) \right]\,, \nonumber \\
e^{2 \hat\phi} &=& g_s^2={R_s^2 \over k^2\,\ell_s^2}=\frac{R^3}{k^3 l_P^3}=\frac{4 \pi}{k^2}\sqrt{2\lambda}\,, \nonumber \\
\tilde{F}_{\mu\nu\lambda\rho} &=& \mp{3 \over 8} k\,g_s\,R_s^2\,\epsilon_{\mu\nu\lambda\rho} = \mp{3 \over 8} \,\frac{R_s^3}{\ell_s}\,\epsilon_{\mu\nu\lambda\rho}\,, \nonumber \\
F_{ab} &=& 2 \,k\,g_s\,J_{ab} = \frac{2 R_s}{\ell_s}\,J_{ab} \,,
\label{ABJbackground}
\eea
where we used  $\epsilon_{\mu\nu\lambda\rho}=\sqrt{-g}\,E_{\mu\nu\lambda\rho}$ with a convention $E_{0123}=-1$ for the numerical totally antisymmetric number $E_{\mu\nu\lambda\rho}$.

In order to simplify notation, we will set up our convention as $R_s=\ell_s=1$ from now on in this paper.

\subsection{Compactification on $\mathbb{CP}_3$}

The compactification on $\mathbb{CP}_3$ yields a theory with $SU(4)\times U(1)$ gauge group and $\mathcal{N}=6$ or $\mathcal{N}=0$ supersymmetry according to the orientation, and this is related to the compactification of the eleven-dimensional supergravity on $S^7$\cite{Nilsson:1984bj}.
The $\mathcal{N}=6$ $SU(4)\times U(1)$ theory obtained from the compactification on $\mathbb{CP}_3$ of the Type IIA supergravity is precisely a truncation of the $\mathcal{N}=8$ $SO(8)$ theory from the compactification on $S^7$ of the eleven-dimensional supergravity.
Here we will review a compactification where the fermions are set to zero in the ground state, which has been studied in \cite{Bak:2010yd}.
The starting point is to solve consistently the Bianchi identities, $d\tilde{F}_{[4]}=-H_{[3]}\wedge F_{[2]}$ and $dH_{[3]}=dF_{[2]}=0$.
We take the ansatz where $\tilde{F}_{\mu\nu\lambda a}=\tilde{F}_{\mu a b c}=H_{\mu\nu a}=H_{a b c}=0\;,\;\;F_{a b}=2J_{a b}\;,\;\;\tilde{F}_{a b c d}=-2\chi\,(J\wedge J)_{a b c d}$ and $\tilde{F}_{\mu\nu a b}=e^{-{\phi\over 2}-\sigma}\,\,\,^\ast\tilde{F}_{\mu\nu}\,J_{a b}$, and then get
\bea
&&H_{\mu a b}=\nabla_{\mu}\chi\,J_{a b} \,\,\,\,\,,\,\,\,\,\, \nabla_\mu (\,e^{-{\phi\over 2}-\sigma} \tilde{F}^{\mu\nu}\,)=
\nabla_\mu\chi\,\,\,^\ast F^{\mu\nu}-{1\over 3}\epsilon^{\nu\mu\lambda\rho}\,H_{\mu\lambda\rho}\,, \nonumber\\
&&\tilde{F}_{\mu\nu\lambda\rho}=-{3\over 8}(\pm1+\chi^2)\, e^{-{\phi\over 2}-9\sigma}\,\epsilon_{\mu\nu\lambda\rho}\,,
\eea
where the four-dimensional hodge dual is then defined by  $\,^*F_{\mu\nu}={1\over 2} \epsilon_{\mu\nu}\,^{\alpha\beta}\,\, F_{\alpha\beta}$ and
$\tilde{F}_{\mu\nu}$ is not a field strength, but simply an antisymmetric tensor field.
The equation for $\tilde{F}_{\mu\nu a b}$ leads to a constraint, which can be solved by introducing $\bar{A}_\mu$ as
$\bar{F}_{\mu\nu}\equiv \partial_\mu \bar{A}_\nu-\partial_\nu \bar{A}_\mu=\tilde{F}_{\mu\nu}-2\chi\,
e^{-{\phi\over 2}-\sigma}\,\,^\ast \tilde{F}_{\mu\nu}-\chi^2F_{\mu\nu}$.
The equation for $H_{\mu\nu\lambda}$ is then solved by $H_{\mu\nu\lambda}={3\over 2}e^{\phi-6\sigma}\, \epsilon_{\mu\nu\lambda\rho}\,
\big(\pm A^\rho-\bar{A}^\rho-\nabla^\rho\psi\,\big)$.
Now with the above results we can perform a consistent, $SU(4)$ invariant dimensional reduction by taking the following metric ansatz,
\bea
\rmd s^2 &=& {1 \over 4} e^{-3\sigma}\rmd s^2_4 + e^{\sigma} \rmd s^2 (\mathbb{CP}_3)\,.
\label{cp3ansatz}
\eea
Then the resulting equations of motion can be derived from the Lagrangian\cite{Bak:2010yd},
\bea
\label{action}
\frac{16 \pi G_4 \mathcal{L}}{\sqrt{-g}} &=& \mathcal{R}-{1\over 2}(\nabla\phi)^2
-{6}(\nabla\sigma)^2-{3\over 2}e^{-\phi-2\sigma}\, (\nabla\chi)^2
-e^{{3\phi\over 2}+3\sigma} \, F_{\mu\nu}F^{\mu\nu} \nonumber\\
&& -3e^{-{\phi\over 2}-\sigma} \, \tilde{F}_{\mu\nu}\tilde{F}^{\mu\nu} -18 e^{{\phi}-6\sigma}(\pm A-\bar{A}-\nabla\psi)^2+
12e^{-4\sigma} -{3\over 2} e^{{3\phi\over 2}-5\sigma} \nonumber\\
&&-{9\over 2} e^{-{\phi\over 2}-9\sigma}\,(\pm1+\chi^2)^2 -6\,\chi^2 e^{{\phi\over 2}-7\sigma} +6\chi e^{-\phi-2\sigma}\,\tilde{F}_{\mu\nu}\,^*\tilde{F}^{\mu\nu} \nonumber\\
&&+6\,e^{-{\phi\over 2}-\sigma}\Lambda^{\mu\nu}
\Big(\,\,\tilde{F}_{\mu\nu}-2\chi e^{-{\phi\over 2}-\sigma}\,\,\,^*
\tilde{F}_{\mu\nu}-\chi^2 F_{\mu\nu}-\bar{F}_{\mu\nu}\,\,\Big)  \nonumber\\
&& +6\,\chi \,\tilde{F}^{\mu\nu}\big(\,\,^*{F}_{\mu\nu}+2\,\chi\, e^{-{\phi\over 2}-\sigma}\,F_{\mu\nu}\,\big)-4 \chi^3 \,{F}_{\mu\nu}\,^*{F}^{\mu\nu}\,,
\eea
where $\Lambda^{\mu\nu}$ plays the role of a Lagrange multiplier and $\tilde{F}_{\mu\nu}$ is an auxiliary field.
Thus, the theory involves three bulk scalars and two bulk $U(1)$ gauge fields\cite{Duff:1997qz}.
While the $\chi$ field is already diagonal, the $\phi$ and $\sigma$ fields can be diagonalized by the linear combinations,
\be
\phi_+={\phi+18\sigma\over2\sqrt{7}}\,,\,\,\,\,\phi_-={\sqrt{3}\big(3\phi -2\sigma\big)\over2\sqrt{7}}\,.
\ee
On the other hand, the gauge kinetic terms can be diagonalized by
\be
A^H_\mu={\sqrt{3}\over 2}\big(A_\mu\mp\bar{A}_\mu\big)\,,
\ \ \ \ \ A^B_\mu={1\over 2}\big(\, A_\mu\pm 3\,\bar{A}_\mu\big)\,.
\ee
The massless field $A_\mu^B$ couples to the di-baryon current $J^B$ of the ABJM theory.

\section{Solutions in Einstein-scalar gravity dual to ABJM}

It is already known that the equations of motion of the dimensionally reduced four-dimensional effective Lagrangian (\ref{action}) allow some exact solutions, which include AdS space, AdS-Schwarzschild and AdS-Reissner-Nordstrom black branes\cite{Bak:2010yd}. In this section we will point out that there is another known solution, i.e. domain wall in the low-energy effective theory. There would also exist a hairy black hole coupled to a scalar field, but we will not treat it in this paper.

\subsection{Domain wall solutions in general Einstein-scalar gravity}

Many stationary domain wall solutions that do not preserve any supersymmetry have been shown to allow for first order-equations by the construction of a fake superpotential\cite{Freedman:2003ax,Celi:2004st}.
The domain walls in question are solutions to the following Lagrangian
\be
\mathcal{L}=\frac{\sqrt{-g}}{16 \pi G_D}\bigg[\mathcal{R} - \frac{1}{2}G_{ab}(\phi)g^{\mu\nu}\partial_{\mu}\phi^a \partial_{\nu}\phi^b - V(\phi)\bigg]
\ee
where $V(\phi)$ is the scalar potential.
The metric ansatz for a flat domain wall is
\be
ds^2=e^{2B(r)}\rmd r^2 + e^{2C(r)}\bigg(-\rmd t^2+\rmd x_1^2+\cdots+\rmd x_{D-2}^2\bigg) \;.
\ee
Consistently with the symmetry of this ansatz, the scalar fields depend only on the $r$-coordinate, $\phi^a=\phi^a(r)$.
Now let us suppose that a function $W(\phi)$ exists such that
\be
V(\phi)=8(D-2)^2G^{ab}(\phi)\partial_aW(\phi)\partial_bW(\phi)-4(D-1)(D-2)W(\phi)^2 \;,
\ee
then we can find first-order equations which give correct second-order equations of motion after differentiation.
Every $W(\phi)$ that obeys the above relation gives us a corresponding domain wall solution.
If $W(\phi)$ is a superpotential of some supersymmetric theory, these first-order equations are the standard BPS equations for domain walls.
On the other hand, if $W(\phi)$ is not related to the supersymmetry, the resulting solutions are called fake supersymmetric.

Now we will consider the following action in four dimensions and ansatz of a static solution with a general scalar potential,
\bea
S&=&\frac{1}{16\pi G_4}\int\textrm{d}^{4}x\sqrt{-g}\;\bigg[\mathcal{R}-\frac{1}{2}(\nabla\phi)^2 - V(\phi)\bigg]\;,\\
ds^2&=&-e^{2C(r)}\rmd t^2+e^{2B(r)}\rmd r^2+e^{2K(r)}(\rmd x^2+\rmd y^2)\;.
\eea
Then we get the effective Lagrangian,
\bea
S\!&=&\!\frac{1}{8\pi G_4}\!\int\!\textrm{d}t\,\textrm{d}r\textrm{d}x\textrm{d}y~\mathcal{L}_\textrm{eff}\,, \\
\mathcal{L}_\textrm{eff}&=&e^{C - B + 2 K}\bigg[ 2C^\prime K^\prime + {K^\prime}^2 - \frac{1}{4}{\phi^\prime}^2 - \frac{1}{2}V(\phi)e^{2B}\bigg] \nonumber \\
&&-\bigg[e^{C - B + 2 K}(C^\prime + 2 K^\prime)\bigg]'\;,
\eea
where $'$ denotes $\frac{\textrm{d}}{\textrm{d}r}$.

To derive a domain wall solution, suppose that $V(\phi)$ has the following form for some function $W(\phi)$,
\be
\label{potentialV}
V(\phi) = 2 \bigg[4\bigg(\frac{\textrm{d} W(\phi)}{\textrm{d} \phi}\bigg)^2 - 3 W(\phi)^2\bigg]\;.
\ee
Then, one can derive the second-order equations of motion from the following first-order equations\cite{Freedman:2003ax, Skenderis:1999mm},
\bea
&& C^\prime + 2 K^\prime \pm 3W(\phi)e^{B} = 0\,, \\
&& C^\prime - K^\prime = 0\,, \\
&& \phi^\prime \mp 4\frac{\textrm{d} W(\phi)}{\textrm{d} \phi}e^{B} = 0\;,
\eea
where it is always possible to set $B=0$ using the definition of the coordinate $r$.
Then the domain wall metric is given by
\bea
ds^2&=&\rmd r^2+e^{2C(r)}\bigg(-\rmd t^2+\rmd x^2+\rmd y^2\bigg) \nonumber\\
&=& \frac{\rmd z^2}{z^2~W(\phi)^2}+\frac{1}{z^2}\bigg(-\rmd t^2+\rmd x^2+\rmd y^2\bigg)\;,
\eea
where $z$ is defined to be $z=e^{-C(r)}$.

\subsection{Domain wall solution in single-scalar gravity}

To show the existence of a domain wall solution let us first consider such a reduction that $A_\mu=\bar{A}_\mu=\psi=\tilde{F}_{\mu\nu}=\chi=\phi_-=0$, which is consistent with all the equations of motion. Then the resulting theory of gravitation is a truncation to Einstein gravity coupled to a single neutral scalar $\phi_+$,
\be
S=\frac{1}{16\pi G_4}\int\textrm{d}^{4}x\sqrt{-g}\;\bigg[\mathcal{R}-\frac{1}{2}(\nabla\phi_+)^2 - V(\phi_+)\bigg],
\ee
where
\be
V(\phi_+) = - \frac{21}{2}e^{-3\phi_+/\sqrt{7}} + {9\over 2} e^{-\sqrt{7}\phi_+}\;.
\ee
One can check easily that the scalar potential $V(\phi_+)$ can be expressed as in (\ref{potentialV}) by
\be
W(\phi_+) = \frac{7}{4} e^{-3\phi_+/2\sqrt{7}} - \frac{3}{4} e^{-\sqrt{7}\phi_+/2}\;,
\ee
where we chose the relative minus sign so that eq.(\ref{phi}) holds in the limit of vanishing $\phi_+$.
Thus, by solving the equations,
\bea
&& C^\prime = K^\prime = \mp W(\phi_+)e^{B}\,, \\
&& \phi_+^\prime = \pm 4\;\frac{\textrm{d} W(\phi_+)}{\textrm{d}\phi_+}e^{B}\;, \label{phi}
\eea
we can find a domain wall,
\bea
ds^2&=& \frac{\rmd z^2}{z^2~W(\phi_+)^2}+\frac{1}{z^2}\bigg(-\rmd t^2+\rmd x^2+\rmd y^2\bigg)\;, \\
\phi_+ &=& f^{-1}(z)\;,
\eea
where $f^{-1}(z)$ is the inverse function of $f(z)$ given by
\be
f(z)=\bigg(e^{\sqrt{7}z/2}-e^{3z/2\sqrt{7}}\bigg)^{-1/3}\;.
\ee
This solution is transformed by defining 1+$\mid\rho\mid=e^{5\phi_+/\sqrt{7}}$ into
\bea
ds^2&=& \frac{4\;\rmd\rho^2}{225[(1+\mid\rho\mid)^{7/10}-(1+\mid\rho\mid)^{3/10}]^2} \nonumber \\
&&+[(1+\mid\rho\mid)^{7/10}-(1+\mid\rho\mid)^{3/10}]^{2/3}\bigg(-\rmd t^2+\rmd x^2+\rmd y^2\bigg)\;,
\eea
which was first obtained in \cite{Bremer:1998zp} in the $S^7$ reduction of the eleven-dimensional supergravity.
The solution is reflection symmetric about $\rho=0$, where the horizon is located with $g_{tt}$, $g^{\rho\rho}$ and $\phi_+$ vanishing and the solution approaches AdS$_{4}$.
It deviates more and more from AdS$_{4}$ as $\mid\rho\mid$ increases.
In the limits where $\rho\rightarrow\pm\infty$ the curvature $\mathcal{R}=3[45(1+\mid\rho\mid)^{-7/5} - 42(1+\mid\rho\mid)^{-1} - 35(1+\mid\rho\mid)^{-3/5}]/8$ tends to zero. So the metric is asymptotically locally flat, while the scalar $\phi_+$ approaches infinity as $\rho\rightarrow\pm\infty$.

The above domain wall solution can be readily oxidized to the original ten-dimensional thoery.
It is given in the string frame as follows,
\bea
ds^2&=& \frac{4\;\rmd\tilde{r}^{\;2}}{9\;(1-\tilde{r}^{-4})^2}+\frac{\tilde{r}^{2/3}}{4}(1-\tilde{r}^{-4})^{2/3}\bigg(-\rmd t^2+\rmd x^2+\rmd y^2\bigg)+\tilde{r}^{\;2}\;\rmd s^2(\mathbb{CP}_3)\;,\;\;\;\nonumber \\
\tilde{F}_{\mu\nu\lambda\rho}&=&\mp\frac{3}{8\;\tilde{r}^{14}}\,\epsilon_{\mu\nu\lambda\rho}\;,\;\;\;\;\;e^{\phi}=\tilde{r}\;,
\eea
where we have transformed the coordinate, $\tilde{r}=(1+\mid\rho\mid)^{1/10}$. It can be shown that its near-horizon geometry is given by AdS$_4\times\mathbb{CP}_3$. This configuration is related via $\tilde{r}=(r^6+1)^{1/4}$ to the dimensional reduction on a circle of the M2-brane solution,
\bea
ds^2 &=& (1+r^{-6})^{-2/3}\bigg(-\rmd t^2+\rmd x^2+\rmd y^2\bigg)+(1+r^{-6})^{1/3}\bigg[\rmd r^2+r^2\rmd s^2(\mathbb{S}^7)\bigg] \\ \nonumber
 &=& \tilde{r}^{-2/3}\bigg[\frac{4\;\rmd\tilde{r}^{\;2}}{9\;(1-\tilde{r}^{-4})^2}+\tilde{r}^{2/3}(1-\tilde{r}^{-4})^{2/3}\bigg(-\rmd t^2+\rmd x^2+\rmd y^2\bigg)+\tilde{r}^{\;2}\;\rmd s^2(\mathbb{S}^7)\bigg]
\eea
of the eleven-dimensional supergravity.

The Einstein-scalar gravity with different sign of potential allows a cosmological solution\cite{Skenderis:2006fb, Skenderis:2006jq, Townsend:2007aw}.
The sign reversion of the scalar potential can be achieved via replacing $\phi_+$ by $\phi_+\pm\sqrt{7}\pi i$,
\be
S=\frac{1}{16\pi G_4}\int\textrm{d}^{4}x\sqrt{-g}\;\bigg[\mathcal{R}-\frac{1}{2}(\nabla\phi_+)^2 + V(\phi_+)\bigg]\;.
\ee
Then one can readily obtain a time-dependent solution,
\bea
&&\phi_+=\frac{\sqrt{7}}{5}\ln(1+\mid\tau\mid)\;,\;\;\;\;\;ds^2= -\frac{4\;\rmd\tau^2}{225[(1+\mid\tau\mid)^{7/10}-(1+\mid\tau\mid)^{3/10}]^2} \nonumber \\
&&+[(1+\mid\tau\mid)^{7/10}-(1+\mid\tau\mid)^{3/10}]^{2/3}\bigg(\rmd r^2+\rmd x^2+\rmd y^2\bigg)\;.
\eea

\subsection{Perturbed solution in multi-scalar gravity}

In this subsection let us consider a different reduction such as $A_\mu=\bar{A}_\mu=\psi=\tilde{F}_{\mu\nu}=\chi=0$, which is also consistent with all the equations of motion. Then the resulting theory of gravitation is a truncation to Einstein gravity coupled to two neutral scalars $\phi_+$ and $\phi_-$,
\be
S=\frac{1}{16\pi G_4}\int\textrm{d}^{4}x\sqrt{-g}\;\bigg[\mathcal{R}-\frac{1}{2}(\nabla\phi_+)^2 -\frac{1}{2}(\nabla\phi_-)^2 - V(\phi_+,\phi_-)\bigg],
\ee
where
\be
V(\phi_+,\phi_-) = - 12e^{-3\phi_+/\sqrt{7} + \phi_-/\sqrt{21}} + \frac{3}{2}e^{-3\phi_+/\sqrt{7}+ 8\phi_-/\sqrt{21}} + {9\over 2} e^{-\sqrt{7}\phi_+}\;.
\ee
Although it is very complicated to solve the all equations of motion analytically, we may obtain the behavior of the small $\phi_-$ fluctuation to the linear order in the $\phi_+$ domain wall background. In this case the $\phi_-$ equation becomes
\be
(1+\mid\rho\mid)^{3/5}\partial_\rho\bigg[\{(1+\mid\rho\mid)^{7/10}-(1+\mid\rho\mid)^{3/10}\}^2\partial_\rho\phi_-\bigg] = \frac{16}{225}\phi_-\;,
\ee
which is solved as
\be
\phi_-=C_1[(1+\mid\rho\mid)^{2/5}-1]^{1/3}+C_2[(1+\mid\rho\mid)^{2/5}-1]^{-4/3}=C_1 r^2 + C_2/r^8
\ee
with constants of integration $C_1$ and $C_2$. Thus, $\phi_-$ behaves like $r^2$ near the horizon($r=0$) and like $1/r^8$ near the asymptotic boundary($r\rightarrow\infty$). The energy density functional of the scalar field $\phi_-$ is given by\cite{Bak:2010ry}
\bea
&&\sqrt{-g} \bigg[ -g^{tt} (\partial_t\phi_-)^2 + g^{\rho\rho} (\partial_\rho\phi_-)^2 + g^{xx} (\partial_x\phi_-)^2 + g^{yy} (\partial_y\phi_-)^2 + 2V(\phi_+,\phi_-)-2V(\phi_+)\bigg]\nonumber \\
&&\sim\sqrt{-g} \bigg[ g^{\rho\rho} (\partial_\rho\phi_-)^2 + 4 e^{-3\phi_+/\sqrt{7}}\phi_-^{~2} \bigg]\,,
\eea
which is clearly nonnegative. Thus, the $\phi_+$ vacuum is stable against the $\phi_-$ fluctuation.

A truncation to Einstein gravity coupled to three scalars $\phi_+$, $\phi_-$ and $\chi$ is also possible and consistent,
\be
S=\int\frac{\textrm{d}^{4}x\sqrt{-g}}{16\pi G_4}\;\bigg[\mathcal{R}-\frac{1}{2}(\nabla\phi_+)^2 -\frac{1}{2}(\nabla\phi_-)^2 -\frac{3}{2}e^{-\frac{2\phi_+}{\sqrt{7}}-\frac{4\phi_-}{\sqrt{21}}}(\nabla\chi)^2 - V(\phi_+,\phi_-,\chi)\bigg],
\ee
where
\bea
V(\phi_+,\phi_-,\chi) &=& - 12e^{-3\phi_+/\sqrt{7} + \phi_-/\sqrt{21}} + \frac{3}{2}e^{-3\phi_+/\sqrt{7}+ 8\phi_-/\sqrt{21}} \nonumber \\
&&+ {9\over 2} e^{-\sqrt{7}\phi_+}(\pm1+\chi^2)^2 + 6\chi^2e^{-\frac{5\phi_+}{\sqrt{7}}+\frac{4\phi_-}{\sqrt{21}}}
\eea
The analysis similar to the above for the $\chi$ equation is more interesting. The potential for $\chi$ is always positive for the choice of upper sign, meanwhile it can be negative for lower sign. Thus, in the skew-whiffed theory\cite{Gauntlett:2009zw} where the scalar field $\chi$ becomes tachyonic with the mass squared $-$2, the $\chi$ perturbation may cause an geometric instability.

\section{Discussion}

In this note, we obtained explicitly a domain wall solution of Einstein gravity coupled to single neutral scalar field $\phi_+$, which is a consistent reduction of type IIA supergravity dual to the ABJM model. Since the mass squared of the bulk scalar mode $\phi_+$ is $m^2_{\phi_+}=18$, the dimension of the corresponding dual operator $O_{\phi_+}$ is $\Delta=6$ \cite{Bak:2010yd}. Such an operator can be constructed using the scalars $Y^I$ and the fermions $\Psi_I$ of the ABJM field theory, where the index $I$ runs over 1 to 4 to label the $SU(4)$ (anti-)fundamental representation. The $\Delta=1$ and $2$ cases were treated in \cite{Bak:2010qb}. To find the explicit form of the chiral primary operator of $\Delta=6$ would be interesting.

In this paper, when considering the AdS-Einstein-double-scalar gravity we studied only neutral solutions.
However, it seems possible to generalize them to magnetically charged configurations which satisfy the equations of motion of the following Lagrangian,
\be
\frac{16 \pi G_4 \mathcal{L}}{\sqrt{-g}}=\mathcal{R}-\frac{(\nabla\phi_+)^2}{2} -\frac{(\nabla\phi_-)^2}{2} - \bigg(e^{\frac{3\phi_+}{\sqrt{7}}+\frac{2\sqrt{3}\phi_-}{\sqrt{7}}} + 3e^{-\frac{\phi_+}{\sqrt{7}}-\frac{2\phi_-}{\sqrt{21}}}\bigg)F^{\mu\nu}F_{\mu\nu} - V(\phi_+,\phi_-)\;. \nonumber
\ee
It would be interesting to pursue this possibility.
Although in this paper we did not treat a hairy black hole solution, it is an crucial issue to study.
Once the hairy black hole background is obtained, we may understand better the low energy dynamics of the ABJM system using AdS/CFT correspondence.
Understanding the structure of the higher derivative terms is also important because they provide a lot of information about the unitarity and renormalizability properties of the theory.
The first corrections to the effective action of ten-dimensional type IIA string theory involve eight derivative terms occuring at string tree level and at string one-loop order \cite{Gross:1986iv,Green:1997di,Antoniadis:1997eg,Kiritsis:1997em,Peeters:2000qj,Peeters:2001ub,Giusto:2004xm}.
To compactify the corrected type IIA supergravity to four dimensions and to study its solutions would be interesting.
We leave them for future investigation.

\section*{Acknowledgments}

The author acknowledges stimulating discussions with H. S. Yang. This work was supported by the World Class University grant number R32-10130.

\end{document}